\documentclass[conference]{IEEEtran}
\IEEEoverridecommandlockouts

\usepackage{cite}
\usepackage{amsmath,amssymb,amsfonts}
\usepackage{graphicx}
\usepackage{textcomp}

\usepackage{amsthm,stfloats,booktabs,multirow,bbding,pifont,url,color,makecell,bm}

\usepackage{algorithm}
\usepackage{algpseudocode}
\usepackage{colortbl}  
\usepackage{tabularx}
\usepackage[table,xcdraw]{xcolor}
\usepackage{subcaption}
\usepackage[normalem]{ulem}

\def\BibTeX{{\rm B\kern-.05em{\sc i\kern-.025em b}\kern-.08em
    T\kern-.1667em\lower.7ex\hbox{E}\kern-.125emX}}
\begin{document}

\title{Video-to-Audio Generation with Fine-grained Temporal Semantics}

\author{\IEEEauthorblockN{1\textsuperscript{st} Yuchen Hu}
\IEEEauthorblockA{\textit{Tencent AI Lab} \\
Beijing, China \\
uniychu@tencent.com}
\and 
\IEEEauthorblockN{2\textsuperscript{nd} Yu Gu}
\IEEEauthorblockA{\textit{Tencent AI Lab} \\
Beijing, China \\
colinygu@tencent.com}
\and
\IEEEauthorblockN{3\textsuperscript{rd} Chenxing Li}
\IEEEauthorblockA{\textit{Tencent AI Lab} \\
Beijing, China \\
chenxingli@tencent.com}
\and
\IEEEauthorblockN{4\textsuperscript{th} Rilin Chen}
\IEEEauthorblockA{\textit{Tencent AI Lab} \\
Beijing, China \\
rilinchen@tencent.com}
\and
\IEEEauthorblockN{5\textsuperscript{th} Dong Yu}
\IEEEauthorblockA{\textit{Tencent AI Lab} \\
Seattle, USA \\
dyu@tencent.com}
}

\maketitle

\begin{abstract}
With recent advances of AIGC, video generation have gained a surge of research interest in both academia and industry (e.g., Sora). 
However, it remains a challenge to produce temporally aligned audio to synchronize the generated video, considering the complicated semantic information included in the latter.
In this work, inspired by the recent success of text-to-audio (TTA) generation, we first investigate the video-to-audio (VTA) generation framework based on latent diffusion model (LDM).
Similar to latest pioneering exploration in VTA, our preliminary results also show great potentials of LDM in VTA task, but it still suffers from sub-optimal temporal alignment.
To this end, we propose to enhance the temporal alignment of VTA with frame-level semantic information.
With the recently popular grounding segment anything model (Grounding SAM), we can extract the fine-grained semantics in video frames to enable VTA to produce better-aligned audio signal.
Extensive experiments demonstrate the effectiveness of our system on both objective and subjective evaluation metrics, which shows both better audio quality and fine-grained temporal alignment.\footnote{Demo page is available at: \url{https://sounddemos.github.io/vta-sam/}}
\end{abstract}



\section{Introduction}
\label{sec:intro}
Vision and hearing constitute the main senses of we humans for perceiving the world surrounding us, which are usually complementary to each other and both indispensable~\cite{baltruvsaitis2018multimodal,afouras2018deep,hazarika2020misa,ma2021end,zhu2021deep,shi2022learning}.
Recent advances of AIGC enables neural model to produce vivid video from textual descriptions (e.g., Sora~\cite{sun2024sora}), presenting a great potential of AI in benefiting human life.
However, they lack the capability to generate synchronized audio signal to make the video close to realistic scenarios.
Therefore an effective video-to-audio (VTA) method is expected to have outstanding performance on both audio quality and temporal alignment, in which the generated audio should not only match the contents and semantics of the input video but also align temporally with the given video frames.

Inspired by the success of latent diffusion model (LDM) in text-to-video (TTV) generation~\cite{wang2023modelscope,wu2023tune,blattmann2023align,jeong2024vmc,weng2024art,yang2024cogvideox}, recent works achieve a promising progress in text-to-audio (TTA) generation~\cite{ghosal2023text,blattmann2023align,yang2023diffsound,liu2023audioldm,huang2023make,huang2023make2,ruan2023mm,bai2023accelerating,xue2024auffusion,michaels2024audio,mo2024text,majumder2024tango}.
Despite the good performance, TTA can hardly synchronize the generated audio to target video, due to the lack of temporal semantics in the textual input.
Therefore, latest efforts pioneer the study in direct VTA generation~\cite{luo2024diff,zhang2024foleycrafter,xu2024video} with latent diffusion model.
To illustrate, \cite{xu2024video} investigate various contrastive pre-trained audio/video/text embeddings and data augmentation methods on top of LDM to improve semantic alignment, and \cite{zhang2024foleycrafter} propose a sophisticated temporal adapter to model the audio-video alignment.
Despite the effectiveness, these pioneering studies fail to explicitly model the frame-level video-audio synchronization, which is hard considering the complicated input video representations.

In this work, we propose a fine-grained video-to-audio generation approach with explicit frame-level synchronization.
Specifically, we first investigate the popular latent diffusion model for VTA task, similar to the typical TTA works and latest VTA studies.
To enhance the video-audio synchronization, we propose to leverage the grounding segment anything model (SAM)~\cite{liu2023grounding,kirillov2023segment} to extract detailed object information from each video frame, thus providing fine-grained semantic information to condition the LDM for audio generation.
Extensive experiments demonstrate the effectiveness of our approach in enhancing the temporal semantics, in terms of both objective and subjective evaluation metrics.
Analysis shows the superiority of Grounding SAM in providing fine-grained temporal information in guiding LDM generation.

\begin{figure*}[t]
\centering
\includegraphics[width=0.99\textwidth]{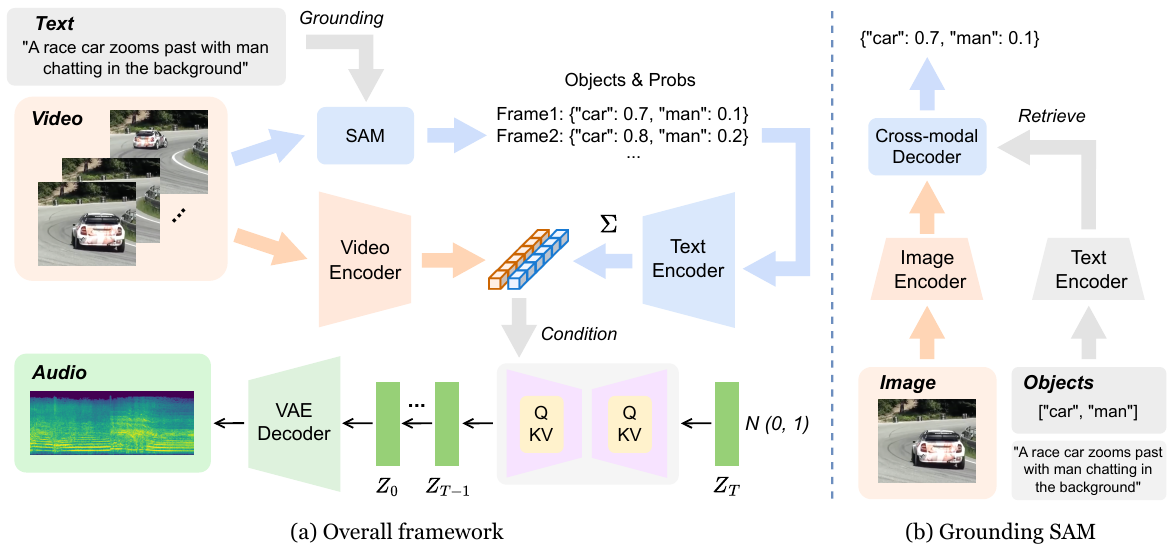}
\caption{The diagrams of (a) overall framework of our VTA approach, (b) Grounding SAM model~\cite{liu2023grounding} from literature.
For each video frame, we use Grounding SAM to extract all the objects and their probabilities to represent fine-grained semantics.}
\label{fig1}
\end{figure*}

\section{Methodology}
\label{sec:method}
\subsection{Overall Framework}
\label{ssec:overall}
Drawing inspirations from the popular TTA works~\cite{ghosal2023text}, we develop a LDM-based VTA framework as shown in Fig.~\ref{fig1}.
It consists of several key components: a video encoder, a Grounding SAM, a Transformer UNet based conditional diffusion model, and an audio mel-spectrum variational auto-encoder (VAE).
Given the input video, we leverage pre-trained CLIP visual encoder~\cite{radford2021learning} to extract visual embeddings.
On the other hand, we introduce Grounding SAM to extract the objects (e.g., ``car'') in each video frame, as well as their appearing probability predicted by SAM.
These information are also transferred into CLIP embedding.
The resulted visual and object embeddings are combined together as the conditioner of LDM for audio generation.
The output representations are sent into VAE decoder to generate mel-spectrum and finally goes into vocoder to produce waveform.

\subsection{Visual Encoder}
In our VTA framework, visual encoder $\mathcal{E}_v$ is used to encode both the semantic and temporal information in video input for comprehensive audio generation.
We leverage the popular CLIP vision encoder in our framework, which is expected to capture the comprehensive visual information including scenes, characters, and events.
Necessary linear projectors are added after it to match the dimension of LDM embeddings. 

\subsection{Grounding Segment Anything Model (SAM)}
The Grounding SAM\footnote{\url{https://github.com/IDEA-Research/GroundingDINO}}~\cite{liu2023grounding} $\mathcal{S}$ is a open-set object detector that can detect arbitrary objects with human inputs such as category names or referring expressions. 
In our VTA framework, for better use of the grounding ability of SAM, we send the paired textual description $t$ to ground and detect the appearing objects from $i$-th video frame $v_i$, e.g., ``car'' and ``man''.
Specifically, we first employ nltk toolkit\footnote{\url{https://www.nltk.org/index.html}} to extract the nouns (i.e., object) $o_i^1, o_i^2, ...$ from input text, thus provides fine-grained prompts for SAM grounding.
Thereafter, SAM consists of two modality encoders to extract image and object embeddings respectively, followed by a cross-modal decoder to retrieve the video frame and predict the appearing probability of each object.
In this way, SAM helps produce comprehensive semantic information of current video frame.

The $j$-th detected object $o_i^j$ is then sent into a CLIP text encoder $\mathcal{E}_t$ to produce the semantic embedding, which are finally weighted summed using the appearing probability $p_i^j$:
\begin{equation}  
\begin{aligned}
  (o_i^1, o_i^2, ...), (p_i^1, p_i^2, ...) = \mathcal{S}(v_i, t), \\
  s_i = \sum_j p_i^j * \mathcal{E}_t(o_i^j),
  \label{eq1}  
\end{aligned}
\end{equation}
$s_i$ is the semantic embedding of $i$-th video frame, and we send $c_o=\{s_i\}_{i=1}^T$ as the final sequential object embedding to enhance diffusion VTA.
It is worth noting that we use the text instead of detected image region to represent the object, which aims to make more comprehensive conditioners for LDM, since the image region is already covered by previous video embeddings.
On the other hand, our incorporation of frame-level object information resolves the limitation of previous TTA methods that relies on utterance-level text.

\subsection{Latent Diffusion Model (LDM)}
Diffusion model is currently among the most popular generative approaches, which consists of diffusion process and reverse process.
Given the original input $x_0$, it first follows a Morkov chain of diffusion steps to add noise to $x_0$ until it finally reaches a pure Guassian distribution, i.e., $x_t \in N (0, 1)$.
Thereafter, during the reverse process, the model learns to denoise the $x_t$ step by step until it recovers the original input $x_0$.
To save the computation cost, latent diffusion model propose to conduct these two process on latent space $Z$ instead of original input space $X$.

In our VTA setting, we have two diffusion conditioners, i.e., $c_v = \mathcal{E}_v(v)$ from video input $v$, where $\mathcal{E}_v$ is CLIP video encoder for feature extraction, and the $c_o=\{s_i\}_{i=1}^T$ from Eq.~\eqref{eq1}.
For step-by-step denoising, we train a neural model $\epsilon_\theta$ to predict the noise with following training objective:
\begin{equation}
  \theta = \arg\min \mathbb{E}_{t\sim U(1, T)} \Vert \epsilon - \epsilon_\theta(Z_t, t, c_v, c_o) \Vert_2^2,
  \label{eq2}
\end{equation}
where we use SAM to extract frame-level semantic information in textual format, and then extract the hidden embedding using CLIP text encoder $\mathcal{E}_t$, which shares the same semantic space with CLIP video encoder $\mathcal{E}_v$.
Here our consideration in utilizing textual output from SAM instead of the image output is that, the visual information is already covered by video embedding.
Therefore, we select the textual output to make comprehensive condition information.

In addition, during reverse process, we adopt the popular classifier-free guidance, which has been proved effective in text-to-audio generation~\cite{ghosal2023text}. Given the latent variable $z_t$ and the two conditioners, we perform the generation both conditionally and unconditionally, and then sum them up with a specified weight $\alpha$.
\begin{equation}
  \epsilon'_\theta = \alpha * \epsilon_\theta(Z_t, t, c_v, c_o) + (1-\alpha) * \epsilon_\theta(Z_t, t),
  \label{eq3}
\end{equation}

\subsection{LDM Conditioned on Video and Objects}
We investigate how to integrate the two conditioners $c_v$ and $c_o$ into diffusion model $\epsilon_\theta$.
Our LDM follows the UNet structure where each module is a Transformer block (i.e., self-attention, cross-attention, feed-forward network).
Denote the latent variable after self-attention module as $z_t$, now we need to incorporate $c_v$ and $c_o$ during cross-attention computation.
Considering that $c_o$ contains more fine-grained temporal information than $c_v$ with the help of SAM, we send the video representations into cross-attention for retrieval and then add the sequential object information to enhance the semantics:
\begin{equation}
  z_t' = \text{Attention}(z_t, c_v, c_v) + c_o,
  \label{eq4}
\end{equation}
We also investigate some other schemes for conditional LDM as discussed in Section.~\ref{ssec:condition_method}.

\subsection{Mel-spectrum VAE and Vocoder}
As introduced before, our diffusion model is performed in latent space, where the variable $z_0$ is extracted by VAE encoder from mel-spectrum input.
Specifically, we use the pre-trained VAE weights from AudioLDM~\cite{liu2023audioldm}.
After generating the target mel-spectrum, we finally send it into Hifi-GAN Vocoder~\cite{kong2020hifi} to obtain the waveform output.

\section{Experiments and Results}
\label{sec:exp_result}

\subsection{Experimental Setup}
\label{ssec:exp_setup}

\noindent\textbf{Dataset.} 
We evaluate the proposed method on public VGGSound dataset~\cite{valentini2016investigating}.
In particular, this dataset contains 550 hours of videos with paired audio-visual events.

\noindent\textbf{Configurations.} 
Our LDM follows the popular Transformer UNet~\cite{ronneberger2015u,petit2021u} structure, which is configured with a cross-attention dimension of 1024 and 8 input and output channels. 
We train our models with learning rate of 3e-5 and no warmup steps, and the batch size is set to 8 per GPU.
All models are trained for a total of 40 epochs on 4 NVIDIA-A100-40GB GPUs. 
During inference stage, we select 1,000 samples from official test set for efficient evaluation.
Specifically, we set the denoising steps to 300, the number of samples per audio to 1, and the guidance scale to 3 for classifier-free guidance.

\noindent\textbf{Metric.} 
For objective evaluation, we use several quantitative metrics, including Frechet distance (FD), Frechet audio distance (FAD), and Kullback–Leibler (KL) to measure the semantic similarity of generated audio with the ground-truth.
Furthermore, we use the AV-Align metric~\cite{yariv2024diverse} to evaluate the temporal alignment of generated audio with input video.
For subjective evaluation, we invite human listeners to evaluate from four different perspectives of perception, i.e., audio quality, semantic alignment, temporal alignment, and the overall quality.
Specifically, we select 30 samples and invite five listeners to evaluate them using a 0-100 score scale~\cite{xu2024video}.

\subsection{Results and Analysis}
\label{ssec:results}

\begin{table}[t]
\caption{Performance comparison with baselines from both TTA and VTA literature, in terms of objective metrics.}
\centering
\resizebox{0.48\textwidth}{!}{%
\begin{tabular}{lcccc}
\toprule[1.2pt]
System & FD $\downarrow$ & FAD $\downarrow$ & KL $\downarrow$ & AV-Align $\uparrow$ \\ \midrule[1.2pt]
\multicolumn{5}{c}{\cellcolor[HTML]{EBEBEB} \emph{Text-to-Audio}} \\
Tango~\cite{majumder2024tango} & 41.7 & 4.7 & 5.5 & 0.173 \\
\multicolumn{5}{c}{\cellcolor[HTML]{EBEBEB} \emph{Video-to-Audio}} \\
Diff-Foley~\cite{luo2024diff} & 30.6 & 4.5 & 5.2 & 0.193 \\
VTA-LDM~\cite{xu2024video} & 30.3 & 4.5 & 5.0 & 0.204 \\
FoleyCrafter~\cite{zhang2024foleycrafter} & 28.3 & 3.8 & 4.8 & 0.233 \\
VTA-SAM (ours) & \textbf{25.1} & \textbf{3.1} & \textbf{4.5} & \textbf{0.255} \\
\bottomrule[1.2pt]
\end{tabular}}
\label{table1}
\end{table}

\begin{table}[t]
\caption{Subjective evaluation results of proposed method and baselines. Larger value means better performance.
}
\centering
\resizebox{0.48\textwidth}{!}{%
\begin{tabular}{lcccc}
\toprule[1.2pt]
\multirow{2}{*}{System} & Audio & Semantic & Temporal & Overall \\
& Quality & Alignment & Alignment & Quality \\ \midrule[1.2pt]
Ground-Truth & 81.7 & 83.8 & 80.5 & 79.9 \\
\multicolumn{5}{c}{\cellcolor[HTML]{EBEBEB} \emph{Text-to-Audio}} \\
Tango~\cite{majumder2024tango} & 72.0 & 56.2 & 52.1 & 57.8 \\
\multicolumn{5}{c}{\cellcolor[HTML]{EBEBEB} \emph{Video-to-Audio}} \\
Diff-Foley~\cite{luo2024diff} & 73.0 & 68.6 & 57.9 & 62.7 \\
VTA-LDM~\cite{xu2024video} & 72.5 & 70.4 & 60.3 & 65.2  \\
FoleyCrafter~\cite{zhang2024foleycrafter} & 74.7 & 70.9 & 61.4 & 65.9 \\
VTA-SAM (ours) & \textbf{75.7} & \textbf{72.4} & \textbf{63.5} & \textbf{66.7} \\
\bottomrule[1.2pt]
\end{tabular}}
\label{table2}
\end{table}

\begin{table}[t]
\caption{Ablation study of input modalities. `V' denotes video, `T' denotes text, `O' denotes the detected objects from Grounding SAM, and `+' denotes combination.}
\centering
\resizebox{0.48\textwidth}{!}{%
\begin{tabular}{lcccc}
\toprule[1.2pt]
Input Modality & FD $\downarrow$ & FAD $\downarrow$ & KL $\downarrow$ & AV-Align $\uparrow$ \\ \midrule[1.2pt]
T-Only & 41.7 & 4.7 & 5.5 & 0.173 \\
V-Only & 30.3 & 4.5 & 5.0 & 0.204 \\
V + T & 28.8 & 3.8 & 4.9 & 0.220 \\
V + O & \textbf{25.1} & 3.1 & \textbf{4.5} & \textbf{0.255} \\
V + O + T & 25.5 & \textbf{3.0} & 4.6 & 0.251 \\
\bottomrule[1.2pt]
\end{tabular}}
\label{table3}
\end{table}

\begin{table}[t]
\caption{Ablation study of conditional LDM. We investigate the optimal method to insert video and object information into LDM as conditioners.
`Z' denotes the latent variable inside LDM, `V' denotes video, `O' denotes the detected objects from Grounding SAM, `@' denotes cross-attention operation, `+' denotes concatenation operation.}
\centering
\resizebox{0.45\textwidth}{!}{%
\begin{tabular}{lcccc}
\toprule[1.2pt]
Method & FD $\downarrow$ & FAD $\downarrow$ & KL $\downarrow$ & AV-Align $\uparrow$ \\ \midrule[1.2pt]
Z @ V & 30.3 & 4.5 & 5.0 & 0.204 \\
Z @ (V + O) & 26.8 & 3.5 & 4.7 & 0.242 \\
Z @ O + V & 28.3 & 3.9 & 4.6 & 0.233 \\
Z @ V + O & \textbf{25.1} & \textbf{3.1} & \textbf{4.5} & \textbf{0.255} \\
\bottomrule[1.2pt]
\end{tabular}}
\label{table4}
\end{table}

\subsubsection{Objective Evaluation}
Table~\ref{table1} illustrates the comparison of our proposed VTA-SAM with previous works.
We use objective metrics for evaluation, including FD, FAD, and KL to measure the semantic similarity of generative audio and ground-truth audio, as well as AV-Align to measure the temporal alignment of generated audio and input video.
We first reproduce Tango, the SOTA text-to-audio approach, to evaluate the TTA performance under our setting.
Compared to that, the first diffusion-based video-to-audio approach Diff-Foley shows significantly better performance in terms of both semantic and temporal alignment, thanks to the much richer information brought by video than text.
Thereafter, the work VTA-LDM investigate the impact of  vision encoders, auxiliary embeddings, and data augmentation techniques on LDM-based VTA approach, which moves one step forward on performance. 
The most recent work FoleyCrafter designs specific semantic and temporal adapters to explicitly improve the objective performance.
In this work, our proposed approach leverages the abundant pre-trained knowledge of grounding SAM to extract fine-grained temporal semantics to improve VTA.
As a result, it presents better performance in terms of all objective metrics, which demonstrates its effectiveness and provides a new insight for VTA task.

\subsubsection{Subjective Evaluation}
Apart from objective metrics, we also conduct subjective evaluation to verify the effectiveness of our approach.
We evaluate from four different perspectives, i.e., 1) audio quality: merely the quality of generated audio, 2) semantic alignment, the consistency of overall semantic information covered by input video and generated audio, 3) temporal alignment, the consistency of frame-level semantic information covered by input video and generated audio, 4) overall quality, the overall quality of generated audio-video pair.
We first observe from Table~\ref{table2} that the ground-truth audio-video pairs enjoy excellent quality (i.e., around 80) from all perspectives, which sets a high upper-bound for VTA task.
In comparison, the TTA baseline performs much worse on these subjective evaluation, except the audio quality metric.
In addition, we also observe that temporal alignment is the most difficult metric (i.e., lowest value in all methods), which exactly matches our major research motivation.
Previous VTA works post some improvements on these four subjective metrics with sophisticated techniques with diffusion model, while the temporal alignment is still sub-optimal (i.e., around 60).
On top of VTA-LDM, our introduction of grounding SAM brings fine-grained semantic information and successfully improves the quality of temporal alignment (i.e., 60.3$\rightarrow$63.5).
However, we are also aware that its absolute performance is still limited when compared to the ground-truth. 
We expect that the next generation of VTA models may depend on large model and abundant data to achieve groundbreaking progress, referring to the recent success of large language models and Sora.

\subsubsection{Effect of Input Modalities}
Table~\ref{table3} presents the ablation study of different combination of input modalities.
From the first two rows, we observe that VTA outperforms TTA by a large margin, thanks to the richer semantic information brought by videos.
Then, we try to combine both modalities and it yields some improvement as indicated by the third row in Table~\ref{table3}.
However, the improvement of `V + T' over `V-Only' is limited since the text description contains little temporal information.
In comparison, our approach introduces grounding SAM to extract the object information in each video frame and weight them using predicted appearing probability and CLIP text encoder.
As a result, we can obtain fine-grained temporal semantics to align the generated audio to input video, which produces significantly better performance in terms of all objective metrics.
In addition, we observe that combine incorporating both fine-grained object information and coarse-grained text information fails to bring further improvement, we speculate that the former is already sufficient to enhance audio generation, where the latter seems to be redundant.

\subsubsection{Effect of Different Conditioning Methods}
\label{ssec:condition_method}
Given the optimal combination of `V + O' for conditioner, Table~\ref{table4} further investigates different methods to incorporate them into LDM.
Inspired by previous work~\cite{xu2024video}, we leverage cross-attention operation, with the latent variable Z as query and the video conditioner `V' as key/value, to fuse V into LDM.
On top of that, we try to incorporate our introduced object conditioner `O' into LDM via three methods, i.e., 1) concatenate V and O as the key/value, 2) leverage O as key/value and add V after cross-attention, 3) leverage V as key/value and add O after cross-attention.
Results indicate that the last scheme shows the best performance.
We speculate that the extracted object information is already fine-grained enough, which makes the cross-attention retrieval unnecessary and thus may explain why simple addition at last yields the best performance.

\section{Conclusion}
\label{sec:conclusion}
In this work, we propose a sophisticated video-to-audio generation approach using latent diffusion model with explicit synchronization.
Specifically, we first investigate the popular latent diffusion model for VTA task, similar to the typical TTA works and latest VTA studies.
To enhance the video-audio synchronization, we propose to leverage the grounding segment anything model to extract detailed object information from video frames, which provides fine-grained semantic information to condition the LDM for audio generation.
Experiments demonstrate the effectiveness of our approach in enhancing the temporal semantics compared to baselines, in terms of both objective and subjective evaluation metrics.
Further analysis indicates the superiority of grounding SAM in providing fine-grained temporal information in guiding LDM generation.

\vfill\pagebreak

\bibliographystyle{IEEEtran}
\bibliography{IEEEexample}

\end{document}